\documentclass[%
 reprint,
 amsmath,amssymb,
 aps,
]{revtex4-1}

\usepackage{graphicx,subcaption}% Include figure files
\usepackage{dcolumn}% Align table columns on decimal point
\usepackage{bm}% bold math
\usepackage{color}
\newcommand{\bk}{{\bf k}}

\begin{document}

\preprint{APS/123-QED}

\title{TDDFT calculations for excitation spectra of III-V ternary alloys}% Force line breaks with \\$In_xGa_{1-x}As$

\author{Zhenhua Ning}

\affiliation{%
 Physics Department, University of Illinois at Urbana-Champaign, 1110 West Green Street, Urbana, Illinois 61801, USA.
}%

\author{Ching-Tarng Liang}%

\author{Yia-Chung Chang}
\email{yiachang@gate.sinica.edu.tw}
\altaffiliation[Also at ]{Physics Department, University of Illinois at Urbana-Champaign and Physics Department, National Cheng-Kung University, Tainan, Taiwan.}
\affiliation{
 Research Center for Applied Sciences, Academia Sinica, Taipei 115, Taiwan% with \\
}%

\date{\today}% It is always \today, today,

\begin{abstract}
%$\rm In_xGa_{1-x}As$
We adopted the time-dependent density functional theory (TDDFT) within the linear augmented Slater-type orbitals (LASTO) basis and the cluster averaging method to compute the excitation spectra of III-V ternary alloys with arbitrary concentration $x$. The TDDFT was carried out with the use of adiabatic meta-generalized gradient approximation (mGGA), which contains the $1/q^2$ singularity in the dynamical exchange-correlation kernel ($f_{XC,00}(\mathbf{q})$) as $q\rightarrow 0$.
We found that by using wave functions obtained in local density approximation (LDA) while using mGGA to compute self-energy correction to the band structures, we can get good overall agreement between theoretical results and experimental data for the excitation spectra.  Thus, our studies provide some insight into the theoretical calculation of optical spectra of semiconductor alloys.
\end{abstract}

\maketitle

%\tableofcontents

\section{\label{intro}Introduction}
III-V ternary alloys are important materials because of their wide applications in various photovoltaic devices. $\rm In_xGa_{1-x}As$ alloys can potentially outperform GaAs in electron transport properties and their room temperature band gaps are particularly well suited for applications in infrared emitting diodes and detectors\cite{Geelhaar1996}. Improved DC current gain, increased mobilities and saturation velocities make $\rm InAs_xP_{1-x}$ a good material for heterojunction bipolar transistors\cite{Averett2003}. Modulation-doped $\rm InAs_xP_{1-x}$ quantum wells show excellent performance in semiconductor lasers\cite{Shimizu2000}. High single-mode yield makes $\rm InAs_xP_{1-x}$ competitive in laser diode applications\cite{Park2004}. Thus, it is highly desirable to have a better understanding of the electronic properties of these alloys. Kim et al.\cite{Kim2003} and Choi et al.\cite{Choi2007} reported  dielectric functions of $\rm In_xGa_{1-x}As$ and $\rm InAs_xP_{1-x}$, respectively for various compositions $x$ including the endpoint values $x=0$ and $x=1$, which can be used as the database for analyzing these alloys with arbitrary composition $x$. Many theoretical methods, such as Bethe-Salpeter equation (BSE) approach\cite{Sham1966,Hanke1979,Hanke1980,Strinati1982,Strinati1984,Onida1995} and time-dependent local density approximation (TDLDA)\cite{Levine1989,Levine1991}  have been used to calculate the excitation spectra of solids. The BSE approach explicitly treats electron-hole interaction (excitonic) effects by solving the two-particle correlation function, which  have been used to calculate optical spectra of bulk semiconductors and achieved good agreement with experiment\cite{Onida1995}. TDLDA focuses on the response of the interacting system to a weak external perturbation and treat the exchange-correlation potential $V_{xc}$ by local density approximation (LDA)\cite{Hedin1971}. The method has been successful in obtaining the excitation spectra of finite systems but not successful for extended systems\cite{Onida1995}.  Another difficulty, which has been known since the early 1980s, is that the basic local-density approximation and its semilocal extensions severely underestimate the band gap[16-22]. A scissor operator $(\triangle E) P_{c\bk}$ must be used to correct the band gap for its application in solids. Here the energy shift $(\triangle E)$ can be obtained either semiempirically\cite{Baraff1984} or by GW calculation \cite{Stinati198082, Hybertsen1986, Godby1988}, and $P_{c\bk}$ is the projection operator applied on conduction bands only. Unfortunately, it is impossible to use the scissor operator in alloys since it is a composition of two or more semiconductors and one cannot determine which band gap to be corrected. It makes the scissor operator method inapplicable to alloys.

The recently emerged meta-generalized gradient approximation(mGGA) \cite{Becke2006,Tran2009,Voorhis1998,Tao2003} can fix the notorious underestimated band gaps caused by LDA\cite{Perdew1981,Perdew1983,Perdew1985,Perdew1986} without consuming large computational resources. It expands the exchange-correlation potential in terms of not only the density, but also the gradient of the density $\nabla n_{\sigma}(\mathbf{r})$, the kinetic energy density  $\tau(\mathbf{r})$ and(or) $\nabla^2 n_{\sigma}(\mathbf{r})$. mGGA developed by Tran and Blaha's(TB09)\cite{Tran2009} shows great improvement in band gaps which are in excellent agreement with experimental results. The mGGA can be used to generate Kohn-Sham(KS) wave functions and eigenenergies with good band gaps for GaAs, InAs, InP, and their alloys. The idea of using time-dependent density functional theory (TDDFT)\cite{Runge1984,Gross1985} with adiabatic mGGA to compute optical spectrum was introduced by Nazarov et al.\cite{Nazarov2011}. They have implemented this approach into the full-potential linearized augmented plane-wave (FLAPW)\cite{Andersen1975} scheme to calculate optical spectrum for bulk Si and Ge with good success. However, FLAPW needs a large number of plane waves as the basis, which makes it computationally expensive to apply to systems with large number of atoms per unit cell. On the other hand, Sharma et al.\cite{Sharma2011,Sharma2014} proposed a "Bootstrap" kernel for TDDFT that determines the long-range correction(LRC) parameter self-consistently and generated good excitation spectra for a wide range of materials, including  the band-gap correction either with a scissor approximation or the LDA+U approach.

In this work we describe the implementation of TDDFT in the full-potential linear augmented-slater-type orbitals (LASTO) scheme \cite{Davenport1984,Davenport1985a,Davenport1985b,Fernand1989} with adiabatic mGGA to compute optical excitation spectra for alloys with the help of the cluster averaging method. The LASTO approach uses a much smaller number of basis functions than FLAPW, which makes it much easier to extend to systems with large unit cells. It is found that by using mGGA to evaluate the exchange-correlation kernel in TDDFT and the self-energy correction to band energies while using the Kohn-Sham wavefunctions calculated in LDA, we can obtain excitation spectra of ternary alloys in very good agreemnet with experimental results.

%Furthermore, to improve the computation efficiency, all calculation will be implemented based on symmetrized-basis %LASTO\cite{Chang2006}.

This paper is organized as follows. In Sec. II we briefly review the TDDFT concepts and formulas on which our calculation is based. In Sec. III we describe how to model the basic super-cell structures needed in our calculation for alloys. In Sec. IV, we apply the TDDFT approach to compute the optical excitation spectra for the family of ternary alloys including $\rm In_xGa_{1-x}As$ and $InAs_xP_{1-x}$. The results are compared to experimental data. Finally, a summary and future outlook are presented to conclude this paper in Sec. V.

\section{\label{theo}Theoretical methods}
Let us consider the linear response of a semiconductor alloy to a weak optical excitation. The response function $\chi$ can be built from the frequency-dependent, dynamical exchange-correlation (XC) kernel $f_{XC}(\mathbf{r},\mathbf{r'},\omega)$ and the noninteracting Kohn-Sham (KS) response function $\chi_{KS}    (\mathbf{r},\mathbf{r'},\omega)$ according to\cite{Gross1985}
    \begin{equation}\label{fxcknl}
        \chi^{-1}(\mathbf{r},\mathbf{r'},\omega)=\chi_{KS}^{-1}
        (\mathbf{r},\mathbf{r'},\omega)-f_{XC}(\mathbf{r},\mathbf{r'},\omega)-\frac{e^2}{|\mathbf{r}-\mathbf{r'}|}\,,
    \end{equation}
where $f_{XC}(\mathbf{r},\mathbf{r'},\omega)$ is defined as
    \begin{equation}\label{fxcknld}
        f_{XC}(\mathbf{r},\mathbf{r'},\omega)=\frac{\delta V_{XC}[n(\mathbf{r},\omega)]}{\delta n(\mathbf{r'},\omega)}\,.
    \end{equation}
 $V_{XC}[n(\mathbf{r},\omega)]$ is the time-dependent XC potential which is a functional of the electron density $n(\mathbf{r},\omega)$. Following the work of Nazarov and  Vignale\cite{Nazarov2011}, we approximate the Fourier transform of the XC kernel $f_{XC}$ by
    \begin{eqnarray}\label{fxcmggaapp}
        f_{\mathbf{G},\mathbf{G'}}^{xc}\approx-\overline{\frac{\partial\epsilon_{xc}}{\partial \tau}}\chi^{-1}_{KS,s}(\mathbf{G},\mathbf{G'})\,,
    \end{eqnarray}
where  $\mathbf{G}$ and $\mathbf{G'}$ denote reciprocal lattice vectors.  $f_{\mathbf{G},\mathbf{G'}}^{xc}$ has the singularity of the type $f^{XC}_{00}(\mathbf{q})\sim 1/q^2 $ as $q\rightarrow 0$ that the traditional approximations do not provide\cite{Reining2002}.
Here $\epsilon_{xc}$ is the exchange-correlation energy density, which depends on the kinetic-energy density $\tau({\bf r})$ in mGGA\cite{Becke2006,Tran2009,Voorhis1998,Tao2003},
and the overbar in $\overline{\frac{\partial\epsilon_{xc}}{\partial \tau}}$ denotes the average over the unit cell.
% Therefore, this XC kernel $f_{XC}$ is expected to describe the long-range interaction better.
The Fourier transform of the noninteracting KS response function $\chi_{KS}(\mathbf{r},\mathbf{r'},\omega)$ can be expressed in terms of the Kohn-Sham  Bloch states and eigenenergies as
    \begin{widetext}
    \begin{eqnarray}\label{dielkai}
    \chi^{KS}_{\mathbf{G},\mathbf{G'}}(\mathbf{q},\omega)&=&
    \sum_{\nu,\nu',\sigma}\frac{f_{\nu,\mathbf{k}}-f_{\nu',\mathbf{k}+\mathbf{q}}}{\omega-E_{\nu',\mathbf{k}}+ E_{\nu,\mathbf{k}}+i\eta}
    \langle
    \Psi_{\nu,\sigma,\mathbf{k}}(\mathbf{r})| e^{-i(\mathbf{G}+\mathbf{q})\cdot
    \mathbf{r}}|
    \Psi_{\nu',\sigma,\mathbf{k}}(\mathbf{r})\rangle \langle
    \Psi_{\nu',\sigma,\mathbf{k}}(\mathbf{r'})|
    e^{i(\mathbf{G'}+\mathbf{q})\cdot \mathbf{r}}|
    \Psi_{\nu,\sigma,\mathbf{k}}(\mathbf{r'})\rangle\,,
    \end{eqnarray}
    \end{widetext}
where $f_{\nu}$ is the occupation number for the Kohn-Sham  Bloch state $|\Psi_{\nu,\sigma,\mathbf{k}}(\mathbf{r})\rangle$ with quantum number $\nu$ and spin $\sigma$ at wave vector $\mathbf{k}$ (limited in the first Brillouin zone). The KS eigenvalues $E_{\nu,\mathbf{k}}$ and $|\Psi_{\nu,\sigma,\mathbf{k}}(\mathbf{r})\rangle$ are obtained within mGGA. Note that the KS response function adopted in Eq.~\eqref{fxcmggaapp} is the static one, i.e., it does not depend on time or frequency.
The excitonic effect is contained in the macroscopic complex dielectric function $\varepsilon_{M}(\mathbf{q},\omega)$, which is related to the macroscopic average of the response function $\chi$  by
    \begin{equation}\label{dielectric}
        \frac{1}{\varepsilon_{M}(\mathbf{q},\omega)}=1+\frac{4\pi
        e^2}{q^2}\chi_{00}(\mathbf{q},\omega)\,,
    \end{equation}
where $\mathbf{q}$ is the wave vector of a photon and $\chi_{00}$ the  ${\bf G}={\bf G'}=0$ component of the Fourier transform of $\chi(\mathbf{r},\mathbf{r'},\omega)$ .

To evaluate $\varepsilon_{M}(\mathbf{q},\omega)$ in the long-wavelength limit (${\bf q}\rightarrow 0$), we  need to calculate the matrix elements $\langle
    \Psi_{\nu,\sigma,\mathbf{k}}(\mathbf{r})| e^{-i(\mathbf{G}+\mathbf{q})\cdot
    \mathbf{r}}|
    \Psi_{\nu',\sigma,\mathbf{k}}(\mathbf{r})\rangle$ that appear in Eq.~\eqref{dielkai}. If the reciprocal vector $\mathbf{G}$ is nonzero, we can simply set ${\bf q}=0$ and evaluate the matrix elements directly. However, for $\mathbf{G}=0$, the transition matrix element becomes zero at ${\bf q}=0$, and we have to calculate the leading contribution which is linearly proportional to ${\bf q}$. By keeping only the term linear in ${\bf q}$, we obtain the following relation
    \begin{equation}\label{kpidentity1}
    \langle \nu \mathbf{k}\mid e^{-i\mathbf{q}\cdot \mathbf{r}}\mid
    \nu'\mathbf{k}\rangle=\frac{\hbar/m}{E_{\nu',\mathbf{k}}-E_{\nu,\mathbf{k}}}\langle
    \nu \mathbf{k}\mid \mathbf{q}\cdot\mathbf{p}\mid \nu'\mathbf{k}\rangle\,,
    \end{equation}
    where $\mathbf{p}=im[H,\mathbf{r}]/\hbar$ is the momentum operator and $m$ is the electron mass. The calculation of $\lim_{q\rightarrow 0} \varepsilon_{M}(\mathbf{q},\omega)$ by using TDDFT with  $f_{\mathbf{G},\mathbf{G'}}^{xc}$ approximated by Eq.~\eqref{fxcmggaapp} will be referred to as TDDFT-A.

Alternatively, we may  use LDA to calculate the transition matrix elements, which implies keeping the KS wavefunctions in LDA while using mGGA to obtain the self-energy correction to band energies. It has been shown that in GW approximation, it is better to keep the KS wavefunctions obtained in LDA rather than using the wavefuctions obtained in fully self-consistent GW calculation\cite{Hybertsen1986,Aryasetiawan1992}. In this way, the dipole transition matrix elements are given by

    \begin{equation}\label{kpidentity2}
    \langle \nu \mathbf{k}\mid e^{-i\mathbf{q}\cdot \mathbf{r}}\mid
    \nu'\mathbf{k}\rangle^{LDA}=\frac{\hbar/m}{E^{LDA}_{\nu',\mathbf{k}}-E^{LDA}_{\nu,\mathbf{k}}}\langle
    \nu \mathbf{k}\mid \mathbf{q}\cdot\mathbf{p}\mid \nu'\mathbf{k}\rangle^{LDA} \,,
    \end{equation}
Here $\langle
    \nu \mathbf{k}\mid \mathbf{p}\mid \nu'\mathbf{k}\rangle^{LDA}$ denote the LDA momentum matrix elements. Such approximation was also used by Rohlfing and Louie\cite{Rohlfing2000}.
The calculation of $lim_{q\rightarrow 0}\varepsilon_{M}(\mathbf{q},\omega)$ obtained this way will be referred to as TDDFT-B.

\section{Cluster-averaging approach for alloys}
 To calculate the dielectric functions of ternary alloys $\rm A_{x}B_{1-x}C$ (with $x$ varying between 0 and 1), we adopt the cluster-averaging method.  We follow the procedures described in Ref.~\cite{Bernard1987}. Firstly, electronic states of five basic configurations, AC, BC, $\rm A_3BC_4$, $\rm AB_3C_4$ and the AC-BC superlattice,  have to be calculated at the corresponding lattice constants which obey Vegard's law\cite{Vegard1921} with
    \begin{eqnarray}\label{lattcon}
        a_{\rm A_{1-x}B_xC}  = xa_{\rm AB} + (1-x)a_{\rm AC}.
    \end{eqnarray}
%We limit our study to isostructural and isovalent alloys.
The macroscopic  dielectric function of the alloy for a given value of $x$ is calculated via a configuration average with a probability weight of  $P^{(n)}(x)$ for the $n$-th configuration. We have
    \begin{eqnarray}\label{cluster}
        \varepsilon_{M}(\mathbf{q},\omega,x)=\sum_{n=0}^4 P^{(n)}(x)\cdot \varepsilon_{M}({\rm A_{4-n}B_{n}C_4})\,,
    \end{eqnarray}
where  $\varepsilon_{M}({\rm A_{4-n}B_nC_4})$  denotes the macroscopic dielectric function of configuration ${\rm A_{4-n}B_nC_4}$. The probability weights can be calculated by the equation (assuming random distribution)
    \begin{eqnarray}
        P^{(n)}(x)= \left( \begin{array}{c} 4 \\ n \end{array}\right) x^{n}(1-x)^{4-n}\,.
    \end{eqnarray}
 The binary endpoint compounds $AC$ and $BC$ are modeled by zincblende structure with Td symmetry. The remaining configurations are modeled by a $AC-BC$ superlattice and two minority clusters $A_3BC_4$ and $AB_3C_4$.

 Following Ref.~\cite{Bernard1987} we model the $AC-BC$ superlattice by using the primitive tetragonal structure with space group No. 115 in the International Tables for Crystallography or point group $D_{2d}$. It contains 4 atoms per unit cell with the primitive vectors
    \begin{eqnarray}
      & &{\bf a}_1 = (\frac{1}{2}, -\frac{1}{2}, 0)a, \nonumber \\
      & &{\bf a}_2 = (\frac{1}{2}, \frac{1}{2}, 0)a, \\
      & &{\bf a}_3 = (0, 0, 1)a, \nonumber
    \end{eqnarray}
where $a$ is the face-centered cubic lattice constant. Two minority clusters $A_3BC_4$ and $AB_3C_4$ require
a larger unit cell to model them, and we use the primitive cubic structure with space group No. 215 or point group $T_{d}$, which contains 8 atoms in the unit cell with the primitive vectors
    \begin{eqnarray}\label{supercell}
      & &{\bf a}_1 = (1, 0, 0)a \nonumber \\
      & &{\bf a}_2 = (0, 1, 0)a \\
      & &{\bf a}_3 = (0, 0, 1)a. \nonumber
    \end{eqnarray}

In principle, we can use three distinct unit cells for five configurations, the typical zincblende unit cell
for $AC$ and $BC$, the primitive tetragonal structure for AC-BC superlattice, and the primitive cubic structure for $A_3BC_4$ and $AB_3C_4$ supercells. In order to cancel systematic errors (caused by finite sampling in zone integration), we use the largest unit cell among them, the 8-atom supercell specified by Eq.~\eqref{supercell}, for  all three supercell configurations. Their constituent atoms are allowed to relax until they reach equilibrium positions. In general, all atoms can move independently, but we restrict their movements in a way preserving the symmetry of the atoms in their unrelaxed (ideal) positions in the crystal. Note that We do not use the largest unit cell for the bulk configurations $AC$ and $BC$, since we found the results of using the supercell and the bulk unit cell are identical, due to that they have the same Td symmetry and sampling points in zone integration are equivalent.

\section{Results and Discussion}
\subsection{Band structures}
\begin{figure}[!h]
        \centering
        \includegraphics[scale=0.3]{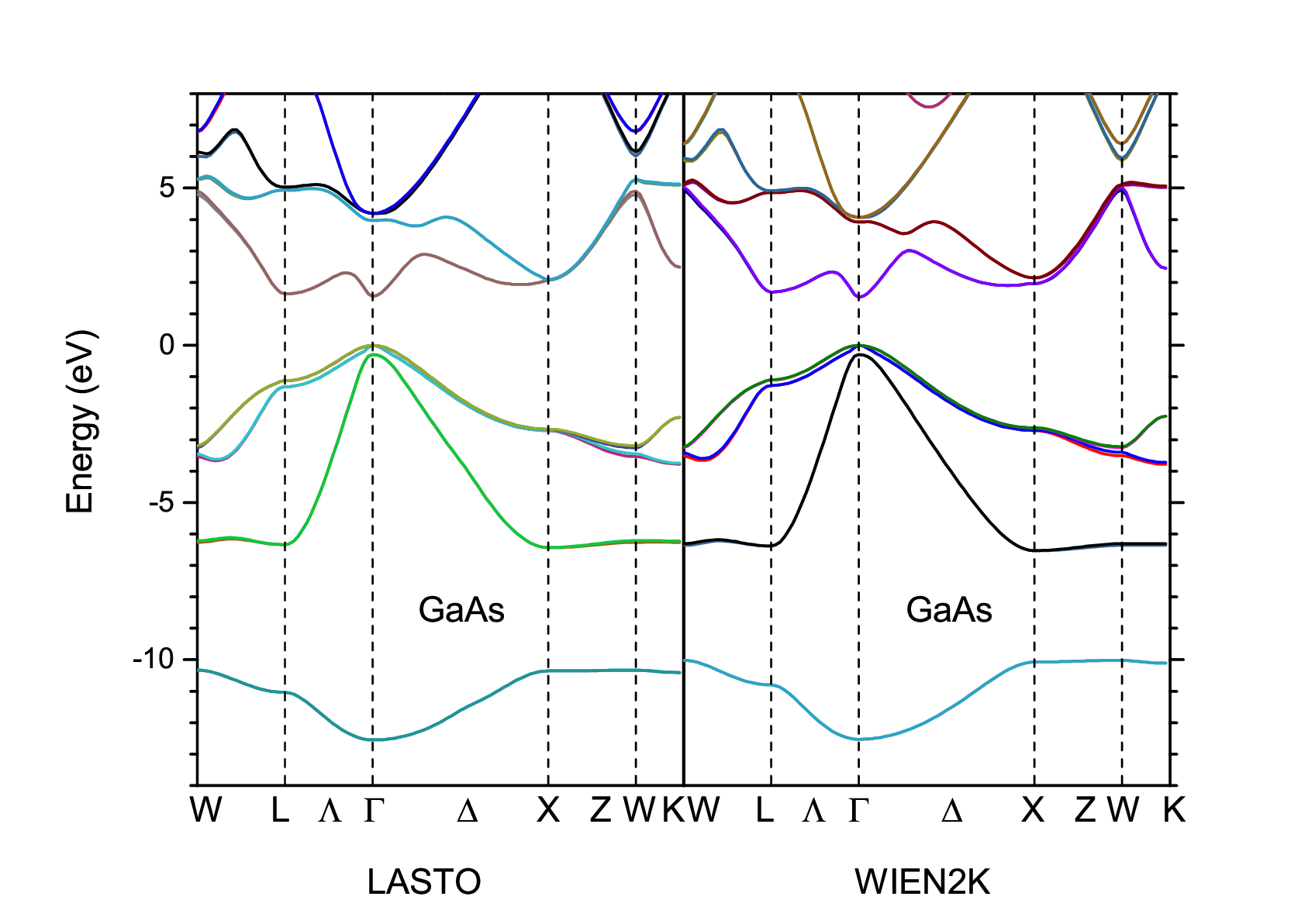} %{GaAs_bands_lasto_wien2k.eps}
        \caption{Band structures of GaAs obtained by LASTO (left) and WIEN2k (right).}\label{gaasband}
    \end{figure}

\begin{figure}[!h]
        \centering
        \includegraphics[scale=0.3]{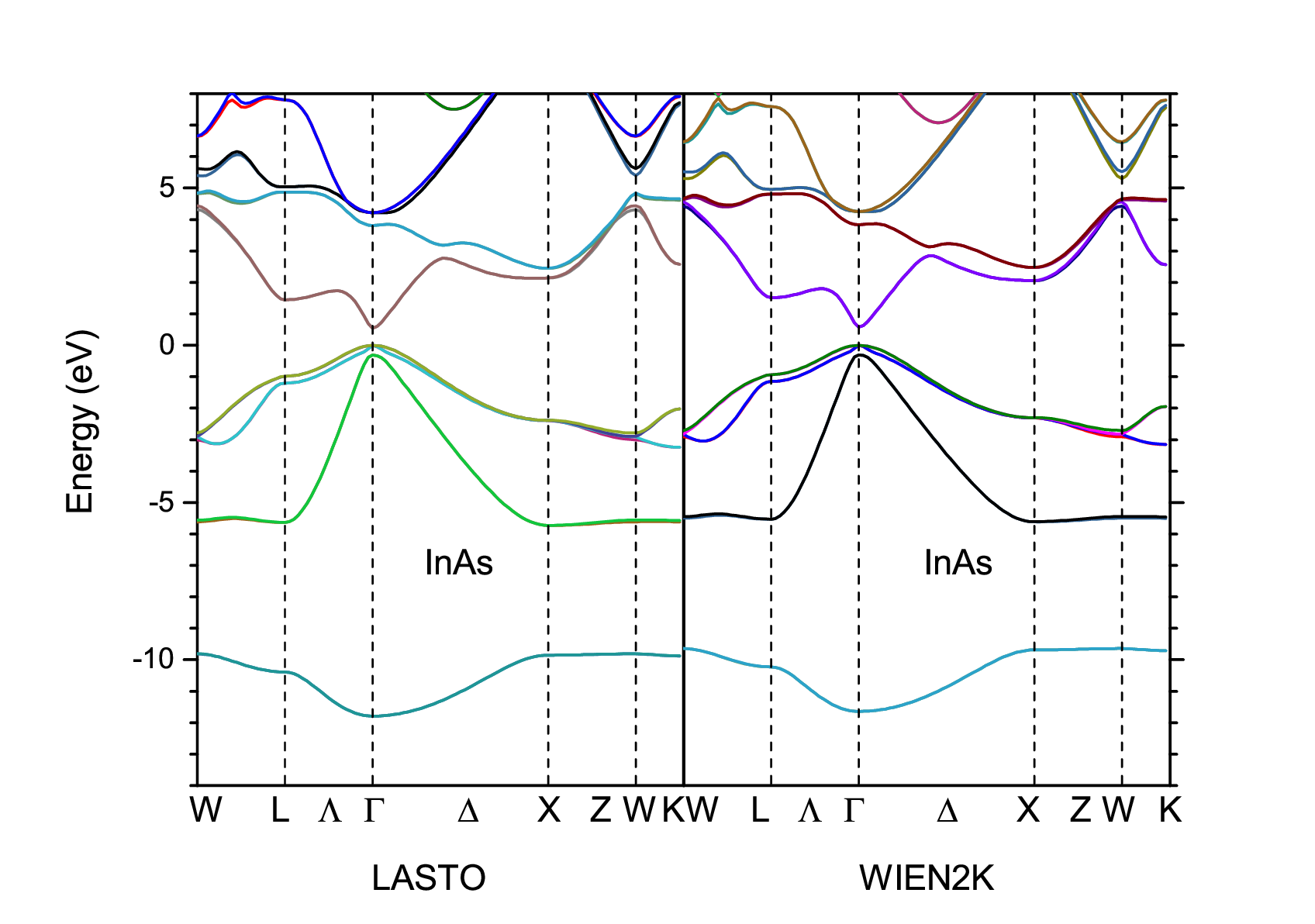} %{InAs_bands_lasto_wien2k.eps}
        \caption{Band structures of InAs obtained by LASTO (left) and WIEN2k (right).}\label{inasband}
    \end{figure}

\begin{figure}[!h]
        \centering
        \includegraphics[scale=0.3]{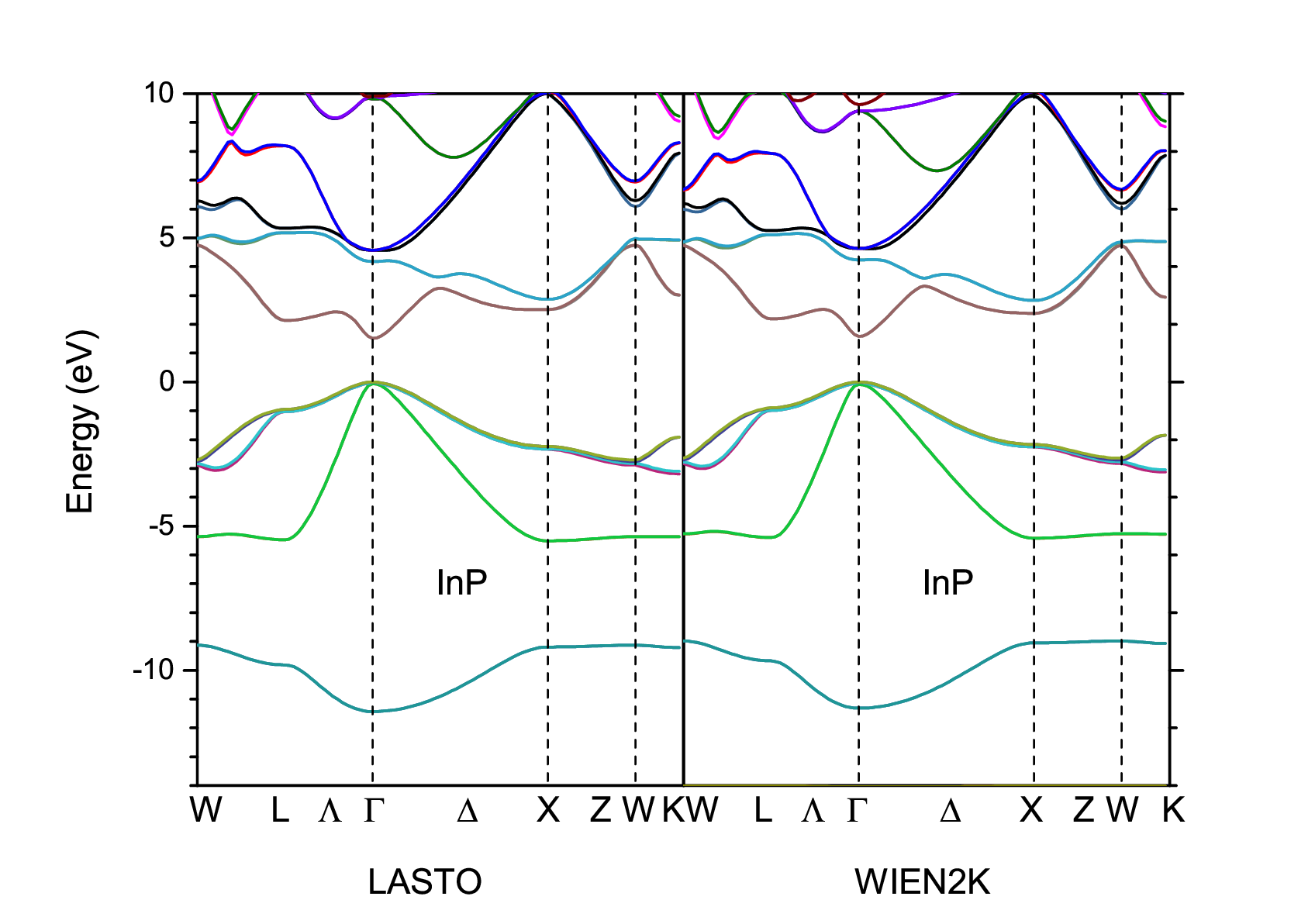} %{InP_bands_lasto_wien2k.eps}
        \caption{Band structures of InP obtained by LASTO (left) and WIEN2k (right).}\label{inpband}
    \end{figure}

\begin{figure}[!h]
        \centering
        \includegraphics[scale=0.3]{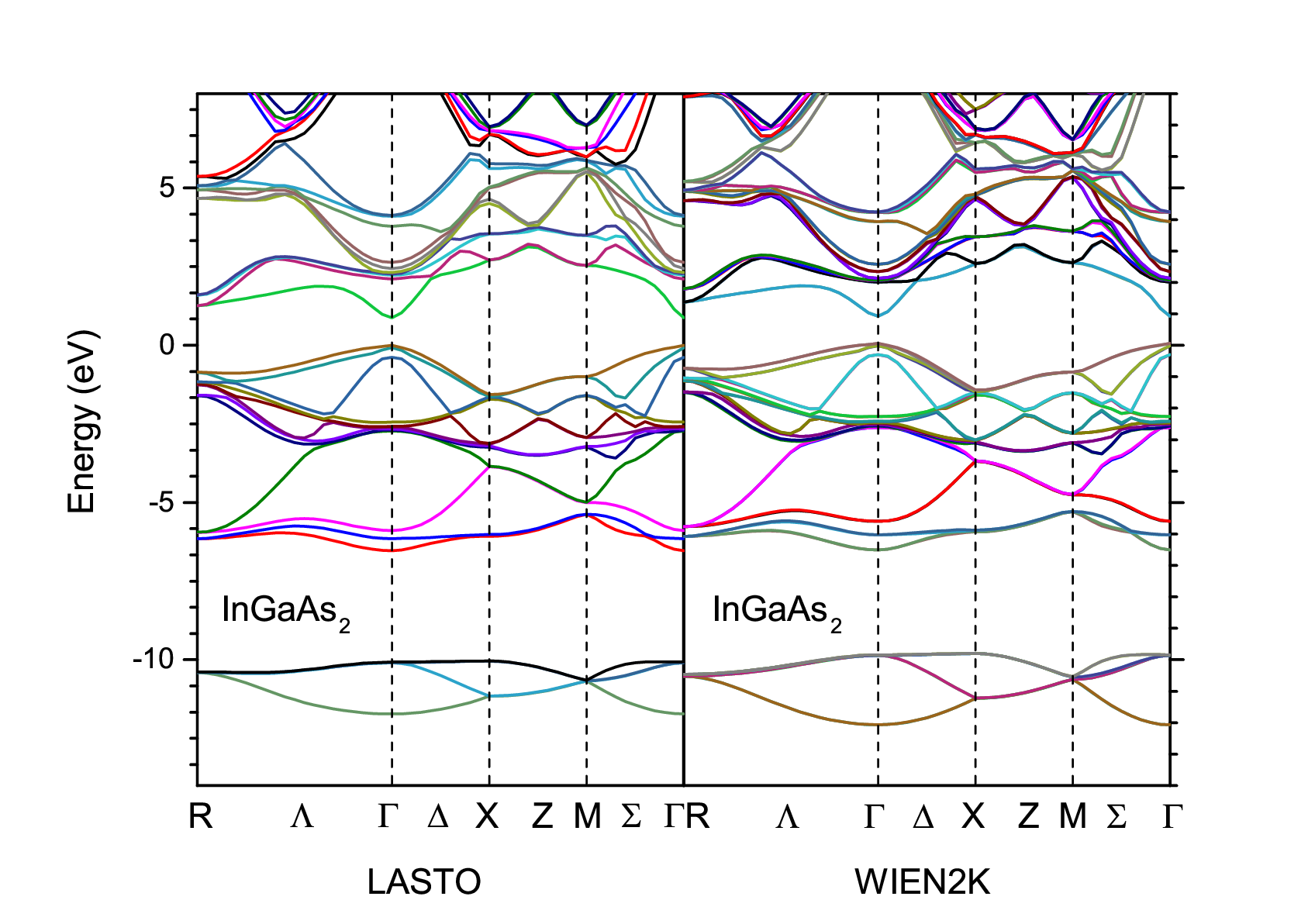} %{InGaAs2_bands_lasto_wien2k.eps}
        \caption{Band structures of $\rm InGaAs_2$ obtained by LASTO (left) and WIEN2k (right).}\label{ingaas2band}
    \end{figure}

\begin{figure}[!h]
        \centering
        \includegraphics[scale=0.3]{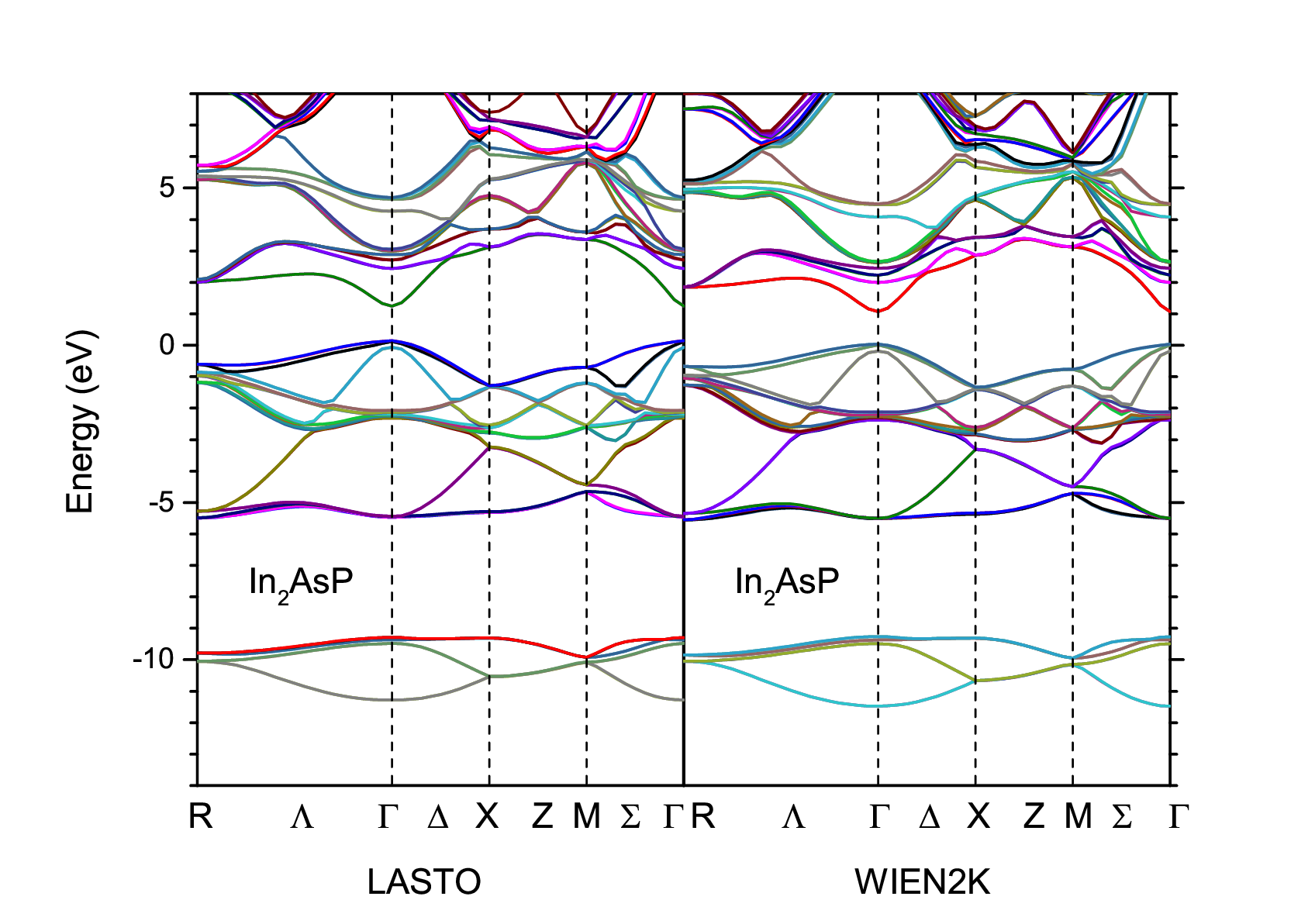} %{In2AsP_bands_lasto_wien2k.eps}
        \caption{Band structures of $\rm In_2AsP$ obtained by LASTO (left) and WIEN2k (right).}\label{in2aspband}
    \end{figure}

The self-consistent KS band structures of constituent materials GaAs (InAs), InGa$_3$As$_4$ (In$_4$As$_3$P), InGaAs$_2$ (In$_2$AsP), In$_3$GaAs$_4$ (In$_4$AsP$_3$), and InAs (InP) for InGaAs (InAsP) alloys are computed  in  mGGA within the LASTO basis by incorporating the TB09\cite{Tran2009} code. The lattice constants of GaAs, InAs, and InP at room temperature are taken from the experimental values compiled in Ref.~\cite{Landolt2006}. For the $AC-BC$ superlattice, $A_3BC_4$, and $AB_3C_4$ supercell structures, their lattice constants are computed by Vegard's law and are listed in Table~\ref{lattconst}.
    \begin{table}
    \caption{The lattice constants (in atomic units) for supercell structures cosidered in modeling the InGaAs and InAsP ternary alloys.}\label{lattconst}
    \begin{tabular}{c|c|c|c|c|c}
    \hline\hline
      $InGaAs_2$&  $InGa_3As_4$& $In_3GaAs_4$ &$In_2AsP$&  $In_4As_3P$& $In_4AsP_3$ \\\hline
      11.07      &    10.87          & 11.26 & 11.27 & 11.36 & 11.18 \\
    \hline\hline
    \end{tabular}
    \end{table}
 We have exploited the  point group symmetry to reduce the computational effort. The relaxation of atoms within the supercell is determined by using WIEN2K\cite{Blaha2001}, which is more reliable than LASTO in terms of structure energy minimization, because it uses the linearized augmented plane wave (LAPW) basis, which is a more flexible basis set than LASTO.  The band structures (including spin-orbit interaction) of GaAs, InAs, InP, InGaAs$_2$ and $\rm In_2AsP$ computed by LASTO and WIEN2K are shown in Figs.~\ref{gaasband} - \ref{in2aspband}, for comparison. Due to the small number of basis functions used in LASTO, the  exponents $\zeta$ used in the Slater basis functions ($\phi_{lm}({\bf r})=r^n e^{-\zeta r}Y_{lm}(\hat r)$) need to be properly chosen to give band gaps in close agreement with WIEN2k. We see that the LASTO results with optimized set of exponents are very close to WIEN2k results in all aspects even though the former uses a much smaller basis set. It is worth noting that the band gaps obtained by the two methods agree within 0.1 eV. For the 8-atom supercell case, the computation time needed to obtain the band structures and KS wavefunctions in LASTO is about a factor 1/8 (1/2) of that for WIEN2k calculation with (without) spin-orbit interaction, indicating the advantage of LASTO over WIEN2k for applications when a large number of KS states are needed in the calculation.


\subsection{Optical excitation spectra}
\begin{figure}[!h]
        \centering
        \includegraphics[scale=0.3]{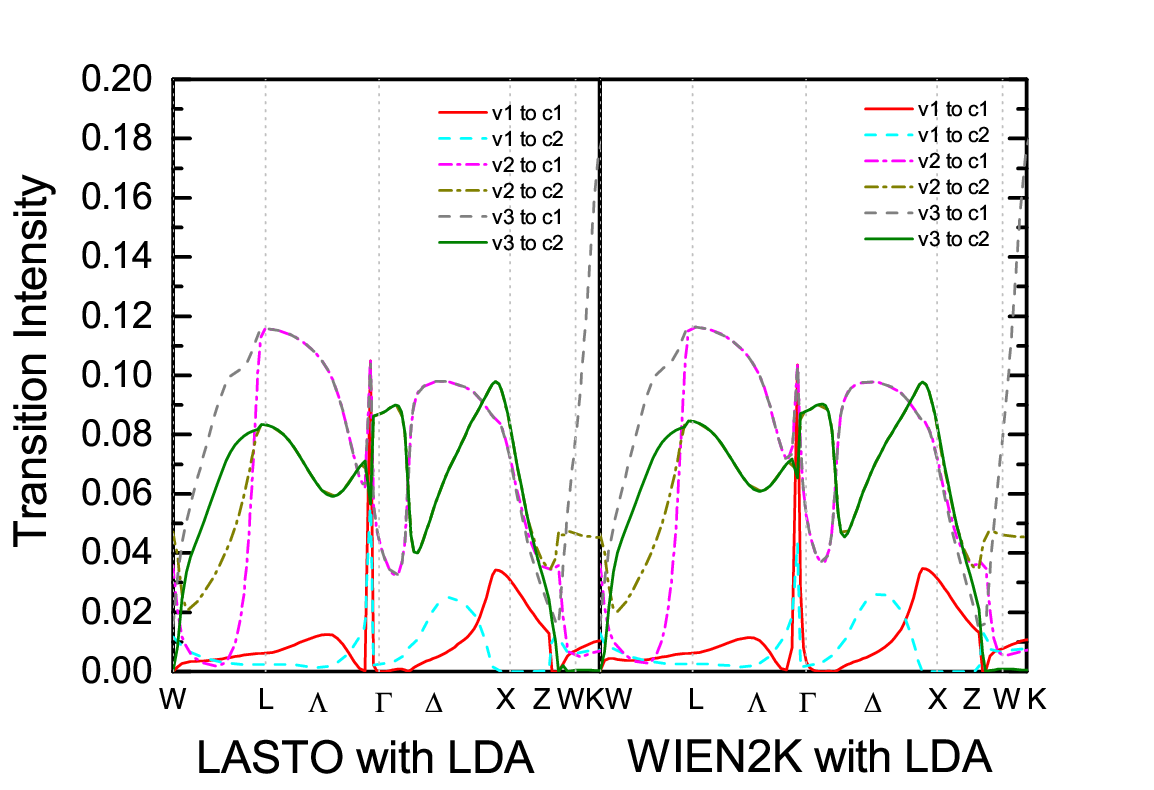} %{GaAs_tran_lda.eps}
        \caption{Squared optical matrix elements of $\rm GaAs$ obtained by LASTO (left) and WIEN2k (right) within LDA.}\label{GaAstrans_lda}
    \end{figure}

\begin{figure}[!h]
        \centering
        \includegraphics[scale=0.3]{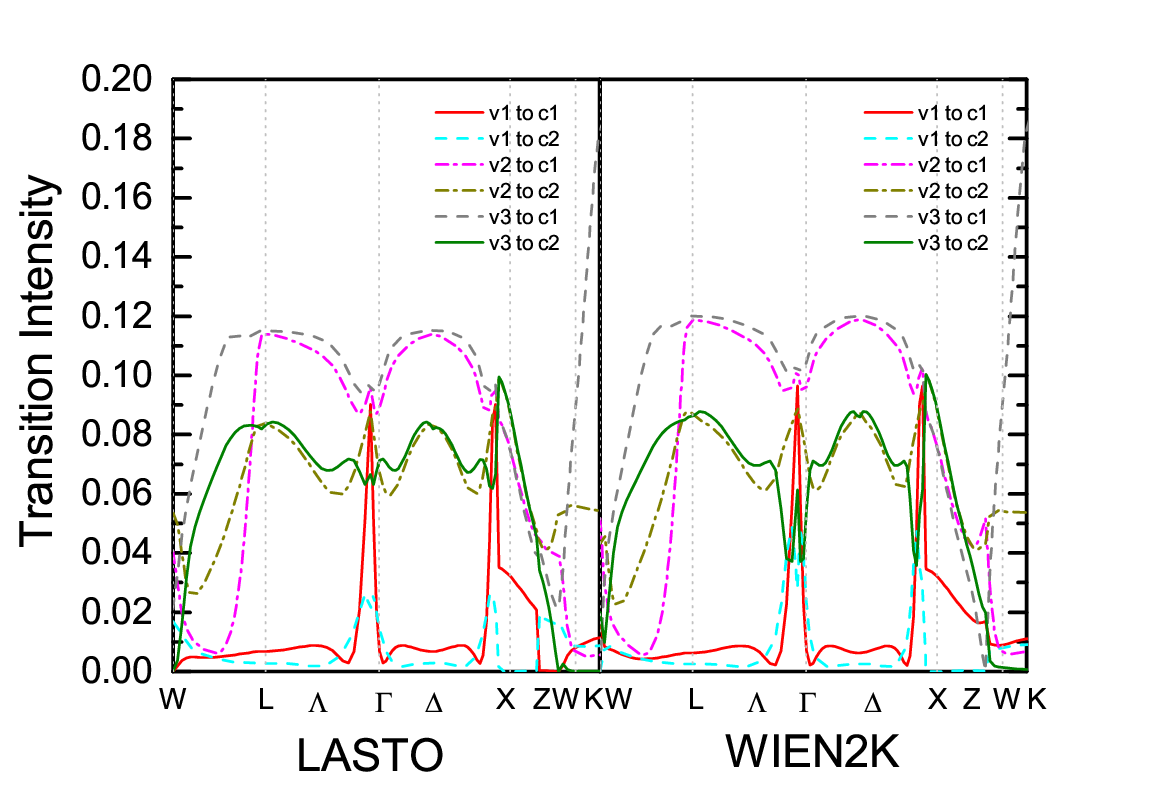} %{GaAs_trans.eps}
        \caption{Squared optical matrix elements of $\rm GaAs$ obtained by LASTO (left) and WIEN2k (right) within mGGA.}\label{GaAstrans}
    \end{figure}

 Using the electronic states obtained with the LASTO basis, we calculate optical excitation spectra for five basic structures through TDDFT. Two methods (TDDFT-A and TDDFT-B) for calculating the excitation spectra have been considered as described in Sec.\ref{theo}. In TDDFT-A, both band structures and wavefunctions are calculated within mGGA. In  TDDFT-B, the mGGA is used to obtain self-energy corrections in the band energies, while the dipole transition matrix elements are evaluated in LDA according to Eq.~\eqref{kpidentity2}.  To check the accuracy of calculated momentum matrix elements in our LASTO code,  we compare the calculated results of $|\langle \nu \mathbf{k}\mid \mathbf{p}\mid \nu'\mathbf{k}\rangle|^2 $) for optical transitions from the highest three valence bands (labeled v1-v3) to the lowest two conduction bands (labeled c1 and c2) obtained both by LASTO and WIEN2k (without including the spin-orbit interaction) in Fig.~\ref{GaAstrans_lda} (within LDA) and Fig.~\ref{GaAstrans} (within mGGA). Due to the possible random mixing of states of degenerate bands in numerical calculations, we take linear combinations of degenerate states to obtain states of fixed symmetry types, which lead to smooth behavior of optical matrix elements for wave-vector along symmetry axes. The comparison shows that results obtained by LASTO and WEIN2k are essentially the same with very minor differences caused by the limited number of orbitals used in the LASTO approach. Similar agreement between LASTO and WIEN2k results is also found for optical transition matrix elements involving higher conduction bands (c3 and c4), although they are not shown here.

\begin{figure}[!h]
    \includegraphics[scale=0.4]{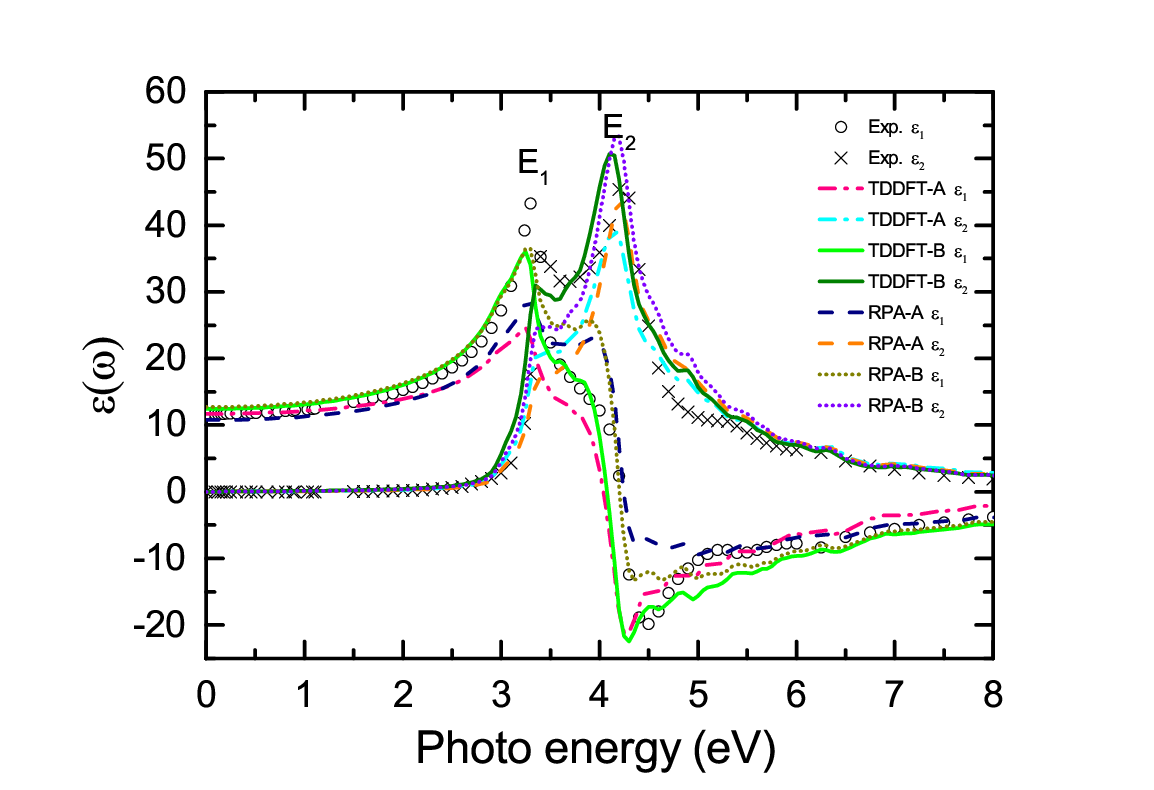} %{Si_five_hybrid_pure.eps}
    \caption{Comparison of optical spectra by various methods for Si. Dash-dotted lines: with pure mGGA (TDDFT-A). Solid lines: with LDA optical matrix elements and self-energy correction in mGGA (TDDFT-B). Dashed lines: RPA-A. Dotted lines: RPA-B. Symbols: Experimental data taken from Ref. \cite{Aspnes1983}. }\label{Si}
    \end{figure}

 %The cutoffs of the reciprocal lattice vectors were set to 2.3 a.u. for bulk structures and 1.5 for supercells. (This is qmax, Gmax is much larger)
 For calculating the optical excitation spectra, the average value of the derivative of the XC energy with respect to the kinetic energy density in Eq.~\eqref{fxcknld} are computed by Voorhis and Scuseria's XC functional\cite{Voorhis1998}. As a bench-mark test, the resulting optical spectra from TDDFT-A and TDDFT-B are presented in Fig.~\ref{Si} for bulk Si together with results calculated in random-phase approximation (RPA) and experimental results from Ref.\cite{Aspnes1983}.  For zone integration, we have used 12x12x12 k-mesh within the irreducible wedge of the Brillouin zone (IWBZ), as generated by the Monkhorst-Pack method\cite{Monkhorst1976}  with shift $(0.083333,  0.25, 0.416667)\frac{2\pi}{a}$.
  The RPA-A and RPA-B results (which do not include the excitonic effect) are obtained by replacing $\chi_{0,0}(\mathbf{q},\omega)$ in Eq.~\eqref{dielectric} by $\chi^{KS}_{0,0}(\mathbf{q},\omega)$ with dipole matrix elements calculated in mGGA and LDA, respectively. In Fig.~\ref{Si}, it is clearly seen that the RPA-A results underestimate both the E1 peak  (due to L-point van Hove singularity) near 3.3 eV and E2 peak  (due to X-point van Hove singularity) near 4.1 eV of the $\varepsilon_2$ spectrum in comparison with experiment, while RPA-B results overestimate the E2 peak (slightly) but still underestimate the E1 peak due to the neglect of the excitonic effect. The TDDFT-A result (dash-dotted curve) improves only slightly with about 15\% increase of the E1 peak over the RPA-A results.  On the other hand, the TDDFT-B result (solid curve) produces significant increase of the E1 peak, bringing much better agreement with the experimental data. Similar behaviors are found for other semiconductors. We thus conclude that it is better to use the TDDFT-B approach to calculate the optical spectra of semiconductors.

\begin{figure}[!h]
    \includegraphics[scale=0.4]{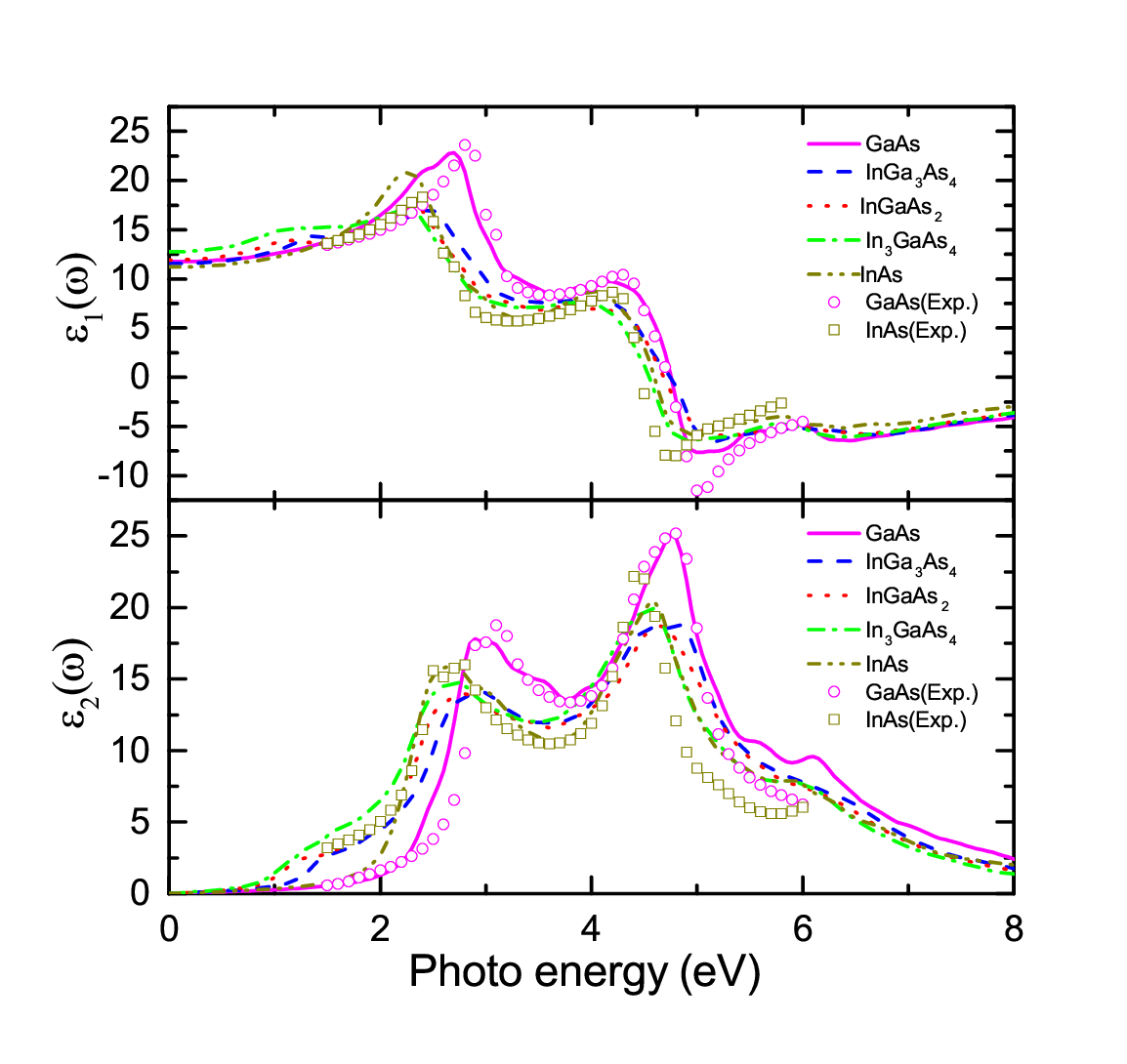} %{InGaAs_five_hybrid_pure.eps}
    \caption{Comparison of optical spectra of five configuration structures for $\rm In_xGa_{1-x}As$ alloy calculated by TDDFT-A approach. Experimental data are taken from Ref.\cite{Kim2003} }\label{ingaas5}
    \end{figure}

\begin{figure}[!h]
    \includegraphics[scale=0.4]{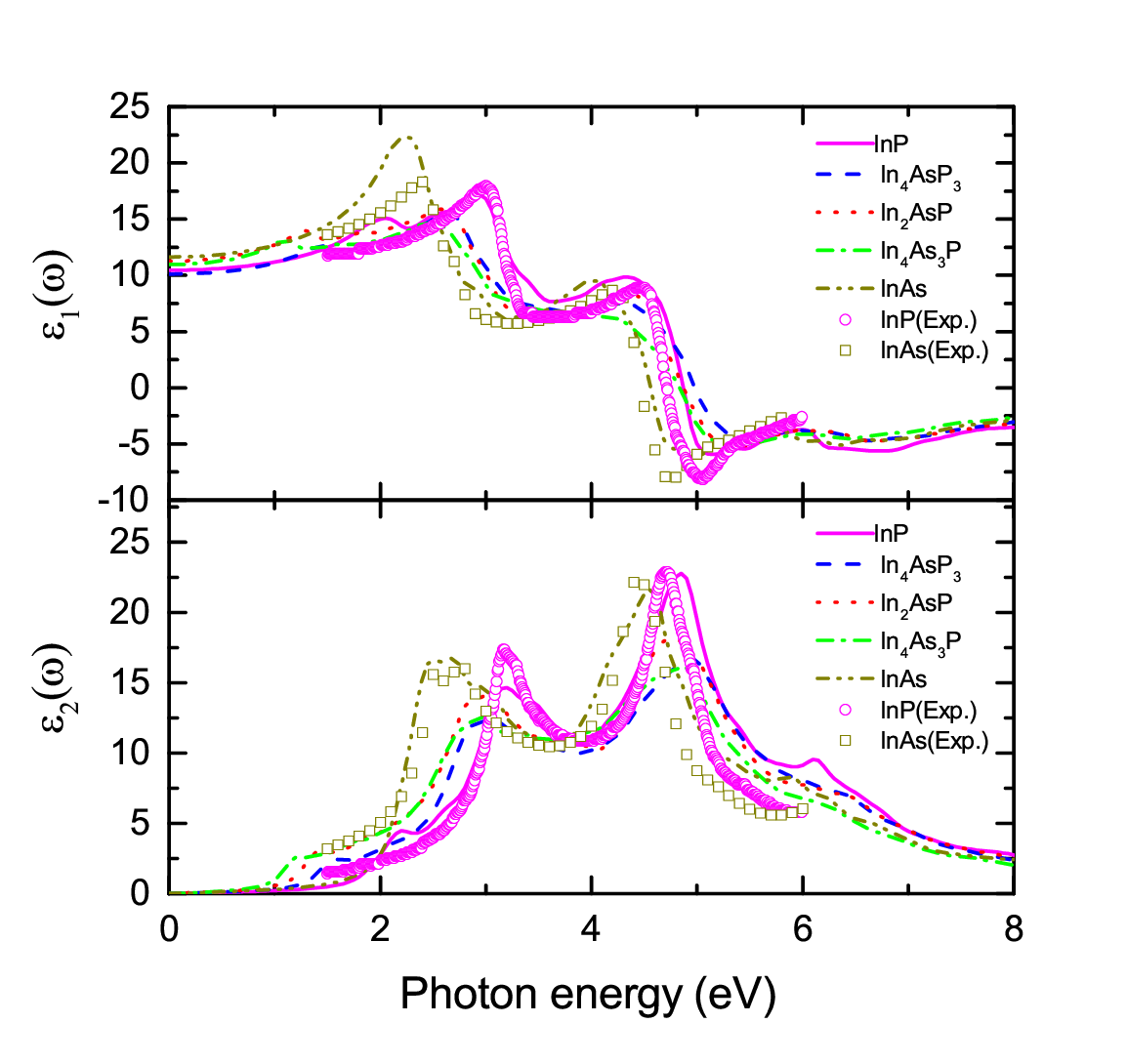} %{InAsP_five_hybrid_pure.eps}
    \caption{Comparison of optical spectra of five  configuration structures for $InAs_xP_{1-x}$ alloy calculated by TDDFT-A approach. Experimental data are taken from Ref.\cite{Choi2007} }\label{inasp5}
    \end{figure}

  Figs. \ref{ingaas5} and \ref{inasp5} show the dielectric functions calculated by TDDFT-A for five configuration structures used in simulating InGaAs and InAsP alloys, respectively. The experimental data for bulk configurations are also included for comparison. We have used a 8x8x8 k-mesh for supercells and 10x10x10 k-mesh for bulk structures within IWBZ to do the zone integration. As seen in Figs. \ref{ingaas5} and \ref{inasp5}, the TDDFT-A results match the experiment data quite well.

    \begin{figure*}
        \centering
        \includegraphics[scale=0.40]{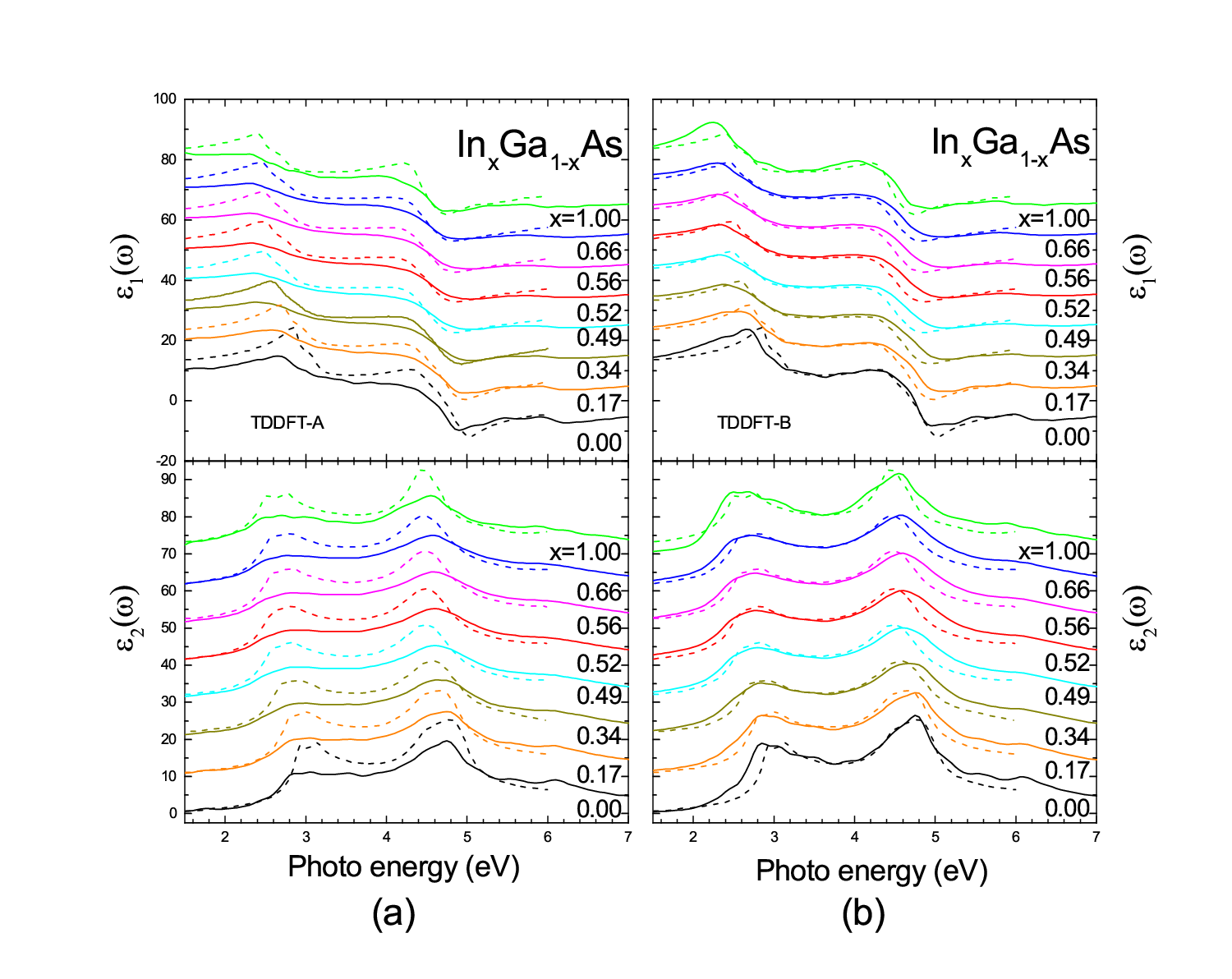}
  \caption{Optical spectra of $\rm In_xGa_{1-x}As$ alloys calculated by (a) TDDFT-A and (b) TDDFT-B. Solid lines are results obtained with cluster averaging method, except for $x=0$ and $x=1$, which correspond to bulk materials. Dashed lines are experimental data from Ref.\cite{Kim2003}. }\label{ingaasspectra}
    \end{figure*}
    \begin{figure*}
        \centering
        \includegraphics[scale=0.40]{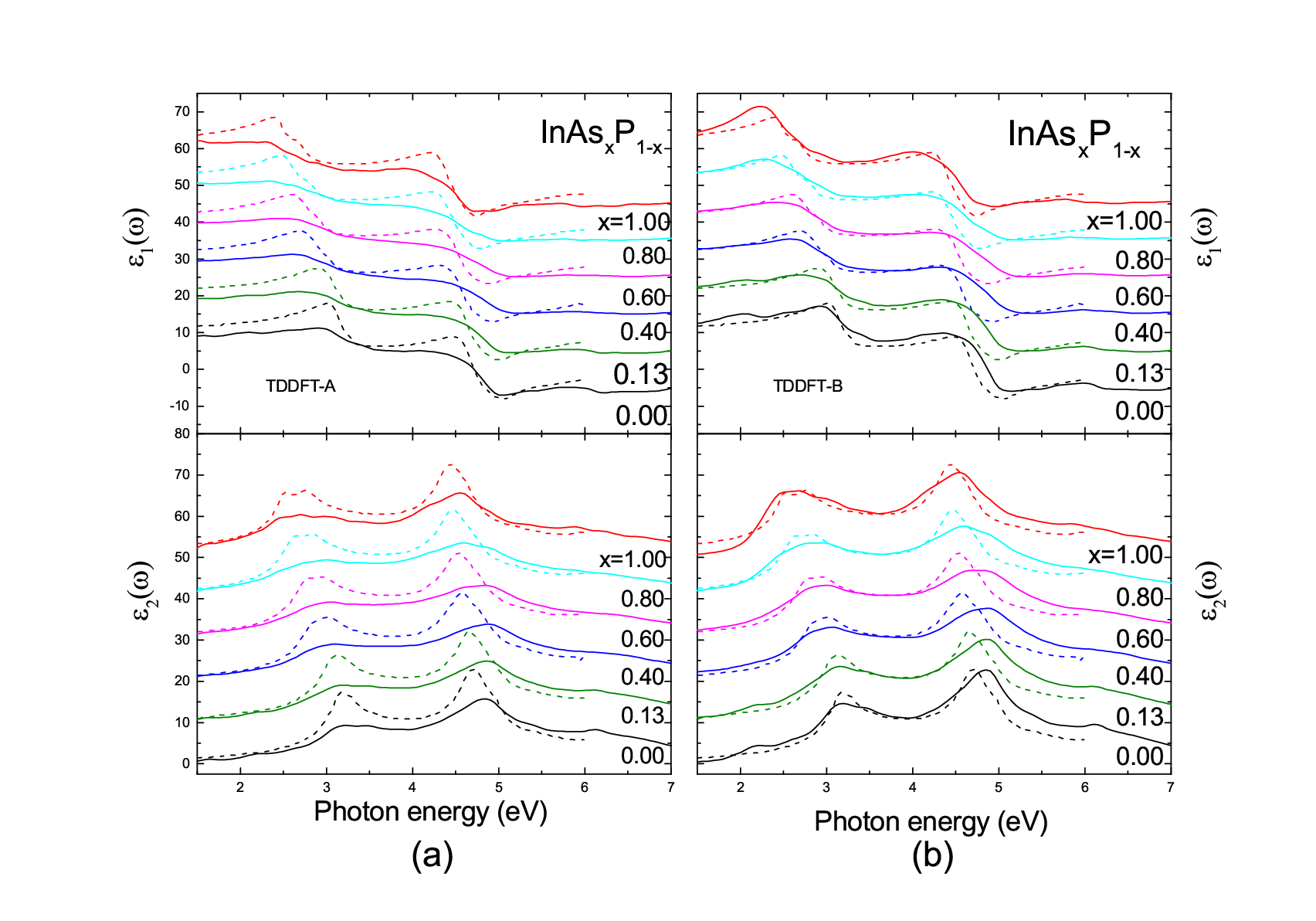}
        \caption{Optical spectra of $\rm In_xGa_{1-x}As$ alloys calculated by (a) TDDFT-A and (b) TDDFT-B. Solid lines are results obtained with cluster averaging method, except for $x=0$ and $x=1$, which correspond to bulk materials. Dashed lines are experimental data from Ref.\cite{Choi2007}. }\label{inaspspectra}
    \end{figure*}
 Next, we adopt the cluster averaging method to calculate the optical spectra of alloys $\rm In_xGa_{1-x}As$ and $\rm InAs_xP_{1-x}$ with arbitrary composition $x$ and compare them with experimental data (dashed lines). The real and imaginary parts of dielectric functions calculated via both TDDFT-A and TDDFT-B methods are presented in Figs.\ref{ingaasspectra} and \ref{inaspspectra}. Solid lines are results obtained with cluster averaging method, except for $x=0$ and $x=1$ which correspond to bulk materials. Dashed lines are experimental data.  The macroscopic dielectric function, $\varepsilon_{M}(\rm A_{4-n}B_nC_4)$  of five configuration structures were computed with TDDFT which includes the many-body interactions through $f_{XC}$ approximated by Eq.~\eqref{fxcknld} within mGGA, and the statistical average based on Eq.~\eqref{cluster} was performed to obtain the macroscopic dielectric functions of the alloys.  It is seen that all spectra calculated by TDDFT-B show a very good match with experimental data. In particular, both the positions and strengths of the $E_1$ (L-point singularity) and $E_2$ (X-point singularity) peaks of $\varepsilon_2$ (imaginary part of $\varepsilon$) spectra agree well with experimental data with differences of height in the range of $0.8\%\sim 4.0\%$ while position deviations in less than $0.3eV$. It is noted that the $\varepsilon_1$ (real part of $\varepsilon$)  spectra are also in very agreement with the experimental data. In contrast, the TDDFT-A approach gives an excitation spectra consistently lower in strength in the entire spectral range, although the energy positions of the E1 and E2 peaks have similar agreement with experiment.  Therefore we conclude that TDDFT-B approach, which use the transition matrix elements calculated in LDA as described by Eq.~\eqref{kpidentity2}, while including the self-energy correction in mGGA gives much better agreement between the theoretical results and experimental data in comparison with the TDDFT-A approach, which adopts the mGGA wavefunctions throughout.  Our studies also illustrate that the cluster averaging method works quite well in obtaining the optical spectra for ternary alloys with statistical average over five basic configuration structures.

\mbox{}

\section{conclusion}
We have used the TDDFT approach combined with mGGA and the cluster averaging method to compute the optical excitation spectra of $\rm In_xGa_{1-x}As$ and $\rm InAs_xP_{1-x}$ alloys with arbitrary composition $x$ and compared them to experimental results with good agreement. This method is simple and efficient. We considered two ways to compute the optical transition matrix elements involved in the calculation of the macroscopic dielectric function: TDDFT-A and TDDFT-B. In TTDFT-A, both the KS wavefunctios and band energies are calculated in mGGA, while in TDDFT-B only the band energies are computed in mGGA thus obtaining the band-gap correction needed to produce excitation spectra with peak positions that match the experiment, while keeping the KS wavefunctions calculated in LDA.   We show that the TDDFT-B approach can generate optical spectra in much better agreement with experiment than the TDDFT-A approach.  This seems to suggest that the KS wavefunctions are more accurately described by LDA, although the band energies need to be corrected by adding the self-energy correction. On the other hand, mGGA is very convenient in providing the self-energy correction to KS band structures and it allows a simple approximation to describe the exchange-correlation kernel $f^{xc}$ in terms of $\overline{\frac{\partial\epsilon_{xc}}{\partial \tau}}$ as given in Eq.~\eqref{fxcmggaapp}, which is needed in the TDDFT calculation to include the excitonic correction to the excitation spectra. We believe this method can be applied in more complicated materials, such as superlattices, quantum wires, and quantum dots in the future.

Since LASTO basis is a much smaller set  than the FLAPW basis used in WIEN2k, it makes TDDFT calculation within LASTO basis efficient enough to handle large supercells (up to 1000 atoms). Thus it can be applied to novel materials and nanostructures with large number of atoms, such as Bucky balls and nanoclusters. Furthermore, the LASTO calculation allows easy extraction of interaction parameters suitable for a tight-binding model, which can be combined with TDDFT to study optical properties of nanostructures with realistic sizes.

\begin{acknowledgments}
This work was supported in part by Ministry of Science and Technology, Taiwan under contract no.  MOST 104-2112-M-001-009-MY2.	We thank V. U. Nazarov for fruitful discussions.
\end{acknowledgments}


\end{document}